\documentclass[conference]{IEEEtran}

\usepackage{cite}

\usepackage{graphicx}
\usepackage{balance}

\usepackage{url}

\begin{document}

\title{ClassCode: \\An Interactive Teaching and Learning Environment for Programming Education in Classrooms
\vspace{-0.5cm}
}


\author{
\IEEEauthorblockN{Ryo Suzuki$^{1, 2}$, Jun Kato$^3$, Koji Yatani$^1$}
\vspace{0.2cm}
\IEEEauthorblockA{$^1$The University of Tokyo, $^2$University of Colorado Boulder,\\ $^3$National Institute of Advanced Industrial Science and Technology (AIST)
}
}

% \author{
% \IEEEauthorblockN{Ryo Suzuki}
% \IEEEauthorblockA{University of Colorado Boulder \\ The University of Tokyo
% }
% \and
% \IEEEauthorblockN{Jun Kato}
% \IEEEauthorblockA{National Institute of Advanced \\ Industrial Science and Technology (AIST)}
% \and
% \IEEEauthorblockN{Koji Yatani}
% \IEEEauthorblockA{The University of Tokyo}
% }

%Telephone: (800) 555--1212, Fax: (888) 555--1212}

\maketitle

\begin{abstract}
Programming education is becoming important as demands on computer literacy and coding skills are growing. Despite the increasing popularity of interactive online learning systems, many programming courses in schools have not changed their teaching format from the conventional classroom setting. We see two research opportunities here. Students may have diverse expertise and experience in programming. Thus, particular content and teaching speed can be disengaging for experienced students or discouraging for novice learners. In a large classroom, instructors cannot oversee the learning progress of each student, and have difficulty matching teaching materials with the comprehension level of individual students. We present ClassCode, a web-based environment tailored to programming education in classrooms. Students can take online tutorials prepared by instructors at their own pace. They can then deepen their understandings by performing interactive coding exercises interleaved within tutorials. ClassCode tracks all interactions by each student, and summarizes them to instructors. This serves as a progress report, facilitating the instructors to provide additional explanations in-situ or revise course materials. Our user evaluation through a small lecture and expert review by instructors and teaching assistants confirm the potential of ClassCode by uncovering how it could address issues in existing programming courses at universities.
\end{abstract}


% \fi
\IEEEpeerreviewmaketitle

\section{Introduction}
Programming education is becoming more important as demands on computer literacy and coding skills are growing. A survey by the United States Labor Department estimates that the number of software developer job positions will increase by 22\% between 2012 and 2022 \cite{handbook2011bureau}. This survey suggests that the number of programming learners will also increase in a similar manner. Therefore, Human-Computer Interaction (HCI) and software engineering research should explore ways to improve programming education to meet future pedagogical demands by society.

Schools and universities still employ a conventional teaching format for programming education; an instructor stands up at a podium and presents teaching materials to students in a classroom. Although many instructors and students are used to this teaching/learning format, we find that it is not suitable for programming education. There are typically a large number of students in programming courses. As computers are commonality nowadays, some students may already have strong coding experience and skills while others are very novice. As a result, the experienced students may find a course disengaging because they already know most of the content. But the same course can be too difficult or its teaching pace is too fast for novice students. Such a wide diversity in technical backgrounds students have makes them easily get disengaged or discouraged.

Instructors are not well supported either in a conventional classroom. In the classroom, students use their terminals to write code, but their working environments typically are not connected or shared with instructors. Thus, it is almost infeasible for them to oversee each student's progress in detail. This leaves instructors unclear about how well students are understating teaching materials during the class.

A recently popular way to learn how to code is online education systems. These systems offer step-by-step tutorials and interactive exercises where learners can deepen their understanding by writing and executing their own codes. This is beneficial from the perspective of programming skill development as they offer an interactive way to quickly experiment what learners have acquired. Although this technology is quite promising, it usually does not consider the use in classrooms, and may not address the issues mentioned above.

\begin{figure}[!b]
\centering
\includegraphics[width=3.5in]{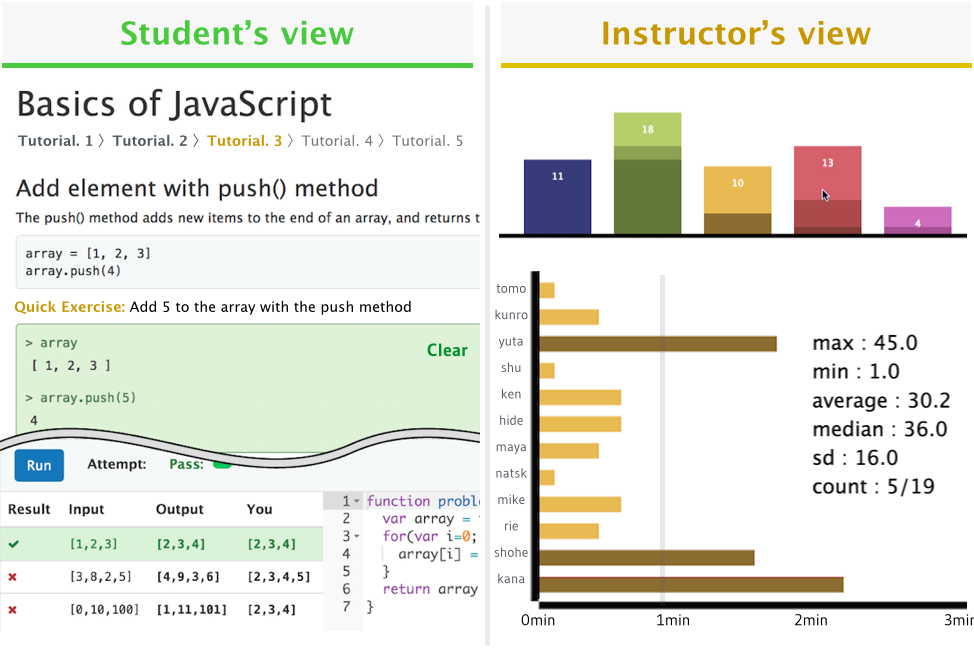}
\caption{Screenshots of the ClassCode interface. ClassCode is designed to support programming education in classrooms.}
\label{fig_1}
\end{figure}

In this paper, we present ClassCode, a web-based learning environment specifically designed for programming education in classrooms (Figure \ref{fig_1}). ClassCode provides the students with an interactive tutorial which allows them to read course materials and perform interactive coding exercises. In this manner, they can quickly study and experiment with the content, effectively forming a closed learning loop. The system also encourages them to share the solutions and comments, facilitating real-time communications in the classroom. ClassCode can track all interactions by students in tutorials, and such data is beneficial for instructors. The system offers instructors an overview of students' performance as well as the detailed view of each learner's activities. In this manner,  instructors can get better informed about their teaching.

The contributions of this work are three-fold: 1) the implementation of ClassCode, an online programming education system designed for classroom use; 2) a user study through a small lecture, confirming the validity of our system design and showing potentials on improving students' learning experience; and 3) expert review supporting the design of ClassCode from instructors' perspective.

\section{Related Work}
As coding is now a fundamental skill required in many academic and industry fields, programming is taught in most universities and schools, recently even in elementary levels. Although the most typical method is perhaps a conventional, one-to-many lecture format, pedagogical studies have investigated different teaching approaches. Pair programming is an approach where two students collaboratively engage in coding \cite{mcdowell2002effects}. This stimulates proactive discussions between them and can lead to better learning than coding by one person. Peer instructions is a teaching method that encourages students to construct their understandings through discussions with peers. They are known to have learning benefits of reducing course failure rates by an average of 61\% \cite{porter2013halving} as well as increasing final exam grades by 5.7\% compared to a conventional teaching method \cite{simon2013we}. These suggest that new ways of teaching and learning can improve the quality of programming education greatly.

Another factor that contributes to reshaping programming education is emerging online education systems. They allow students to learn in a more flexible and interactive way than standard training, such as reading textbooks. We believe that such systems can improve the learning experience of students and help instructors' teaching activities even in classrooms. In this section, we mainly review the literature on online systems for programming education.

\subsection{Online Programming Systems}
There exist a number of online courses that mainly aim to support programming learning. CodeCademy \footnote{\url{http://www.codecademy.com/}} and Khan Academy \footnote{\url{https://www.khanacademy.org/computing/computer-programming}} are great examples of such online services. Courses in these services offer step-by-step tutorials and exercises, and learners can take them as they like. MOOCs (Massive Open Online Courses) is another system enabling online education for programming. It provides visual teaching materials, such as slides and videos, as well as exams like conventional classroom courses.

Online systems can track students' behaviors in detail by recording user interaction. RIMES \cite{kim2015rimes} is a system for authoring, recording, and reviewing interactive multimedia exercises embedded in online lecture videos. It records how students solve problems in videos, audios, and/or drawings while they solve problems. Instructors can then see a quick overview of all students' responses in a thumbnail view. The system also allows the instructor to review a particular student's answer by replaying multimedia records. Their study found that this feature was helpful to identify good answers and misconceptions as well as to understand students' thinking processes. LectureScape \cite{kim2014data} gathers user interaction, such as video clicking and text searching, on online video lectures to improve their learning experience. They discovered that visualizing learners' activities with interaction peaks and word clouds contributed to facilitate them to navigate the video, find specific information, and summarize the contents of the lecture. ASQ \cite{triglianos2013asq} is a web-based presentation tool in which students can directly answer questions given by instructors. This system can be used for hybrid MOOCs, a classroom blended with online lectures. Mudslides \cite{glassman2015mudslide} introduced heat map visualizations for summarizing the understanding levels of students, and found that the system was able to help instructors understand where students were confused during a lecture as well as contributing to improving their teaching skills.

These projects illustrate many benefits of the user interaction tracking capability enabled by online programming education systems. But this has not been well investigated in the context of classroom teaching. One main objective of this ClassCode project is how online programming education systems can enhance learning by students and teaching by instructors.

\subsection{Online Interactive Programming Exercises}
As an exercise is an integral part of online programming education systems, recent research has explored how interactive systems can further improve or augment them. Online Python Tutor \cite{guo2013online} provides an interactive tutorial that helps learners understand how a Python code runs by visualizing changes in variables. Their study revealed the system can help to clarify important concepts in programming, such as function calls and variable scopes. It also suggests that the system can remove the burden of drawing diagrams on a whiteboard in classrooms to visually explain such concepts.
TraceDiff~\cite{suzuki2017exploring, suzuki2017tracediff} leverages this visualization technique to provide personalized and pedagogically useful hints. It automatically debugs student submissions with a program synthesis technique~\cite{rolim2017learning}, then generate personalized feedback about how a student's incorrect code trace diverges from the expected solution, which helps the student locate and identify the bug.
Pex4Fun \cite{tillmann2013teaching} is an interactive teaching and learning platform. This system offers a unique feature of automated grading that enables instructors to grade students' solutions by running a collection of automatically-generated tests. Code Hunt \cite{tillmann2014code, bishop2015code}, is an environment where learners are asked to fill a code for a function that satisfies test cases given by the system. When the learner runs her code, the system automatically creates test cases that match with pre-defined conditions and give results. In this manner, the learner can interactively explore how to write codes.

Debugging is another important aspect of coding, and interactive exercises can facilitate its learning. Lee et al. \cite{lee2014principles} explored a way to teach programming through debugging activities. They first extracted seven design principles for debugging games. They then created a debugging game environment embodying these principles, called Gidget. Their user studies confirmed the benefits of the design principles and Gidget, such as student's learning without dependence to help by instructors.

These explorations suggest that interactive exercises can be engaging and motivating for students. ClassCode is another attempt along this research direction, but our objective is to design them for classroom teaching.

\section{ClassCode Design}
We describe the interface design of the ClassCode system. The ClassCode design is two-fold: from students and instructors. We explain our interface from both points of view.

\subsection{Student's View}
\begin{figure}[t]
\centering
\includegraphics[width=3.5in]{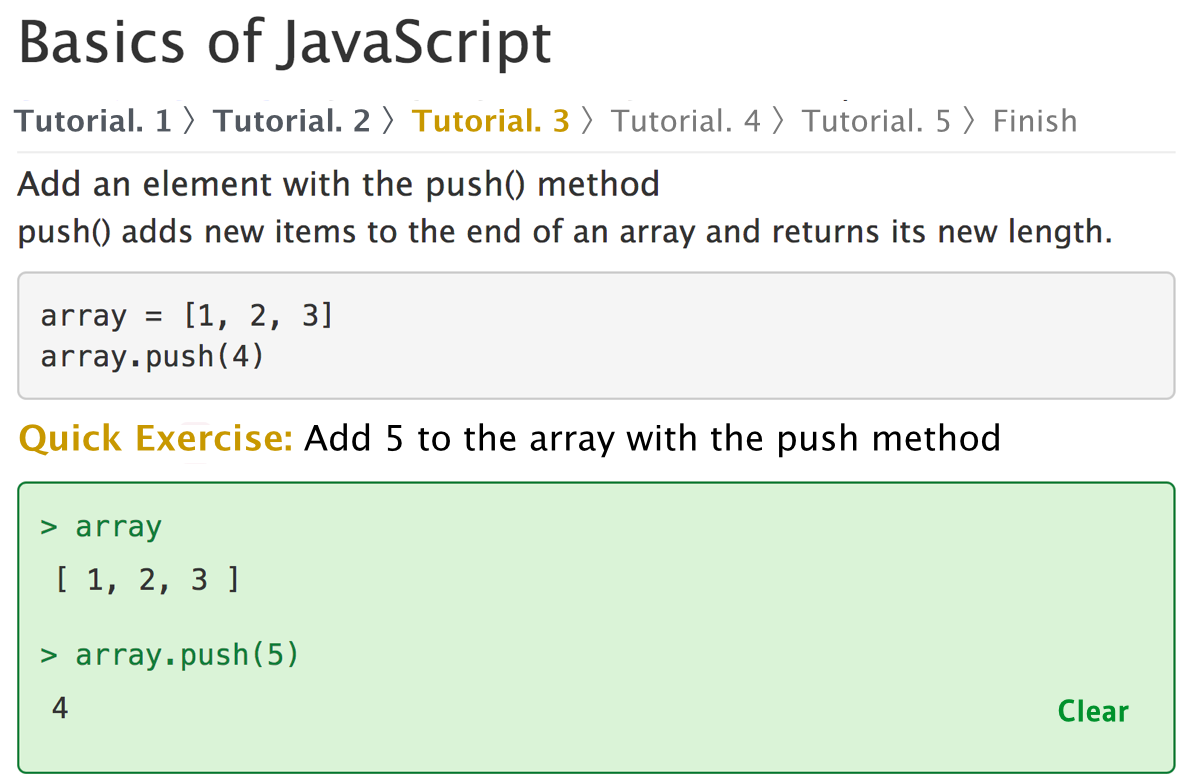}
\caption{A tutorial in ClassCode containing a reading material and quick exercise.}
\label{fig_2}
\end{figure}

When students log into ClassCode, the system shows a tutorial prepared by instructors. As shown in Figure \ref{fig_2}, it contains reading materials (e.g., explanations of functions or algorithms) and short questions, called quick exercises. The students first read the material as instructed in the tutorial. Once they feel comfortable, they can start the exercise associated with the material they have just learned. This design aims to create a closed learning loop of understanding materials and performing quick exercises. When students successfully complete the exercise, the system shows the next reading material. In this way, students can be more confident about their understanding before moving to subsequent parts.

\begin{figure}[t]
\centering
\includegraphics[width=3.5in]{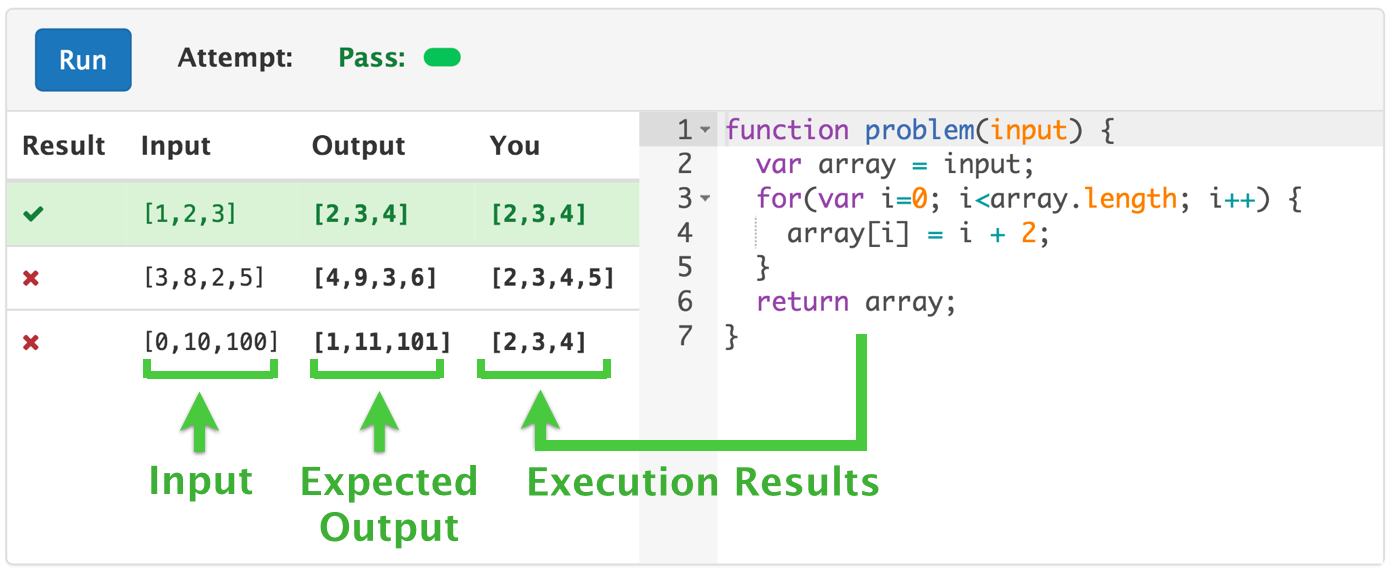}
\caption{A screenshot of a milestone problem interface. Students can write their codes and verify them with test cases prepared by instructors.}
\label{fig_3}
\end{figure}

At the end of a tutorial, the system shows the milestone problem which tests the comprehensive ability to execute what they have learned in the tutorial (Figure \ref{fig_3}). This is inspired by existing coding exercise interfaces like Code Hunt \cite{tillmann2014code}, \cite{bishop2015code} and Pex4Fun \cite{tillmann2013teaching}. It provides a space for writing code as well as a console for interactive debugging. It also shows code test patterns (top-left in the view) given by instructors. Students are asked to write a code such that it passes all test cases. They can easily test their codes by simply pressing the ``Run" button, and the view shows how different the execution results of the test cases are.

\begin{figure}[t]
\centering
\includegraphics[width=3.5in]{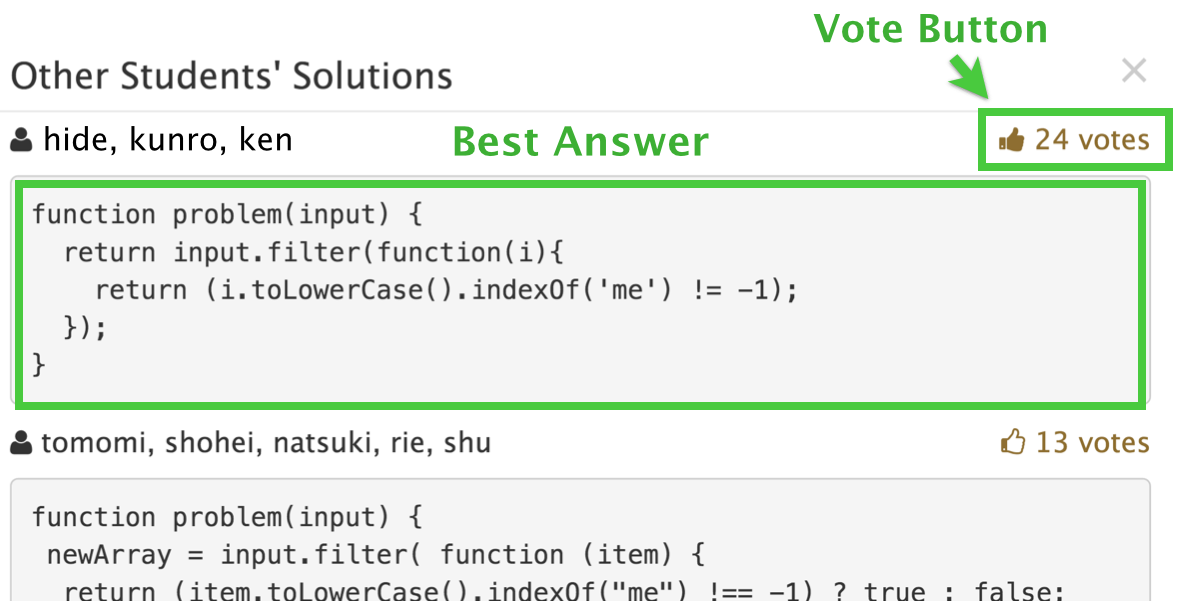}
\caption{At the completion of a milestone problem, students can view others' solutions. This helps them review the problem and solution.}
\label{fig_4}
\end{figure}

When students successfully solve the milestone problem, the system shows other students' solutions (Figure \ref{fig_4}). This view, called a solution gallery, allows students to compare their solutions with others, providing an additional opportunity to review the milestone problem. Students are encouraged to vote for a solution which they agree that is well written. Our current prototype does not do any intelligent clustering like OverCode \cite{glassman2014overcode} and MistakeBrowser~\cite{head2017writing}, but a future system should include it so that students can have a better overview of how their peers have solved the exercise. The students then move to the next tutorial and start learning new materials.

\begin{figure}[t]
\centering
\includegraphics[width=3.5in]{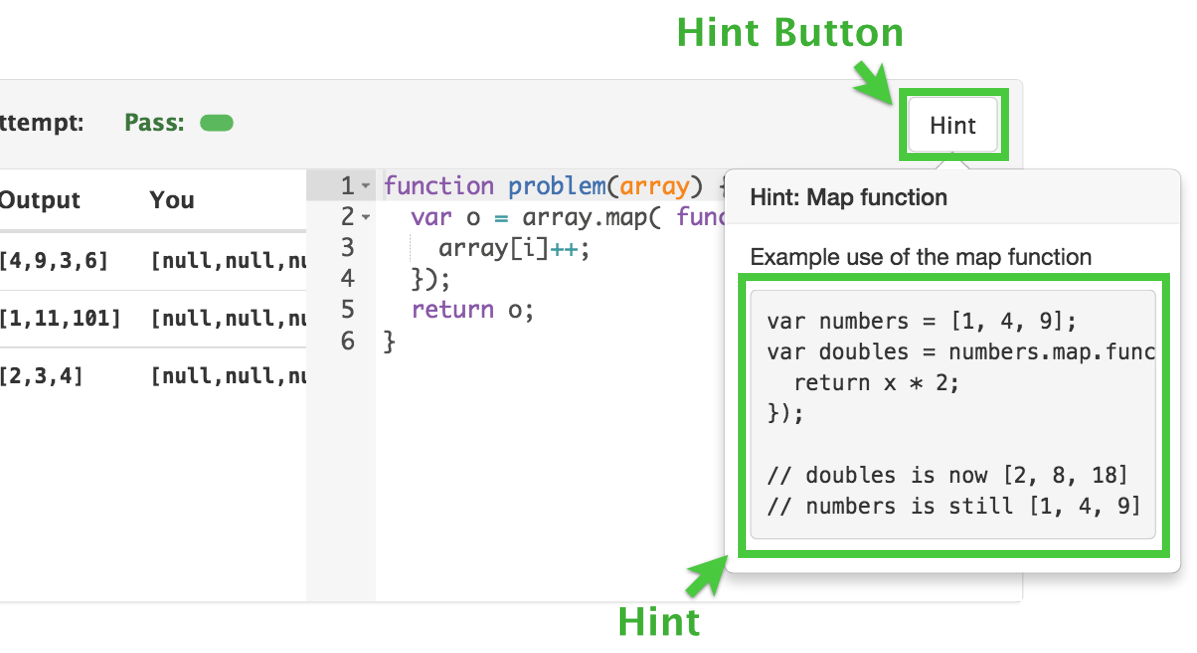}
\caption{Hint button.}
\label{fig_5}
\end{figure}

Some students may need support in solving milestone problems. We provide two ways for students to seek help. When students click the hint button, the system provides additional details to approach to milestone problems (Figure \ref{fig_5}). To prevent students from reading the hint without trying out any code, the button will not appear until they have spent more time than the time threshold set by instructors. When this hint is not sufficient and students have spent even longer than the next threshold, the system enables the help button. When it is clicked, the system shows an online discussion board where students working on the same problem and instructors can read and post comments. They are encouraged to exchange ideas but discouraged to simply share answers by having instructors in conversations. Prior research shows that such an online discussion forum can have a positive impact \cite{coetzee2014chatrooms}, and we aim at a similar effect in ClassCode.

\subsection{Instructor's View}
\begin{figure}[t]
\centering
\includegraphics[width=3.5in]{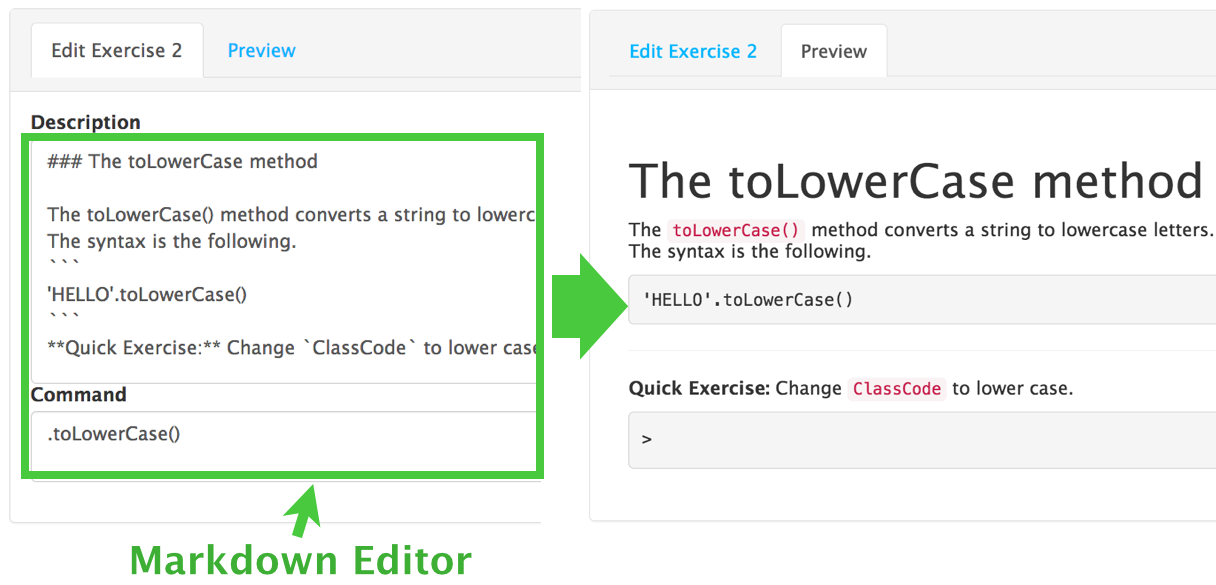}
\caption{Authoring a reading material and quick exercise.}
\label{fig_6}
\end{figure}

\begin{figure}[t]
\centering
\includegraphics[width=3.5in]{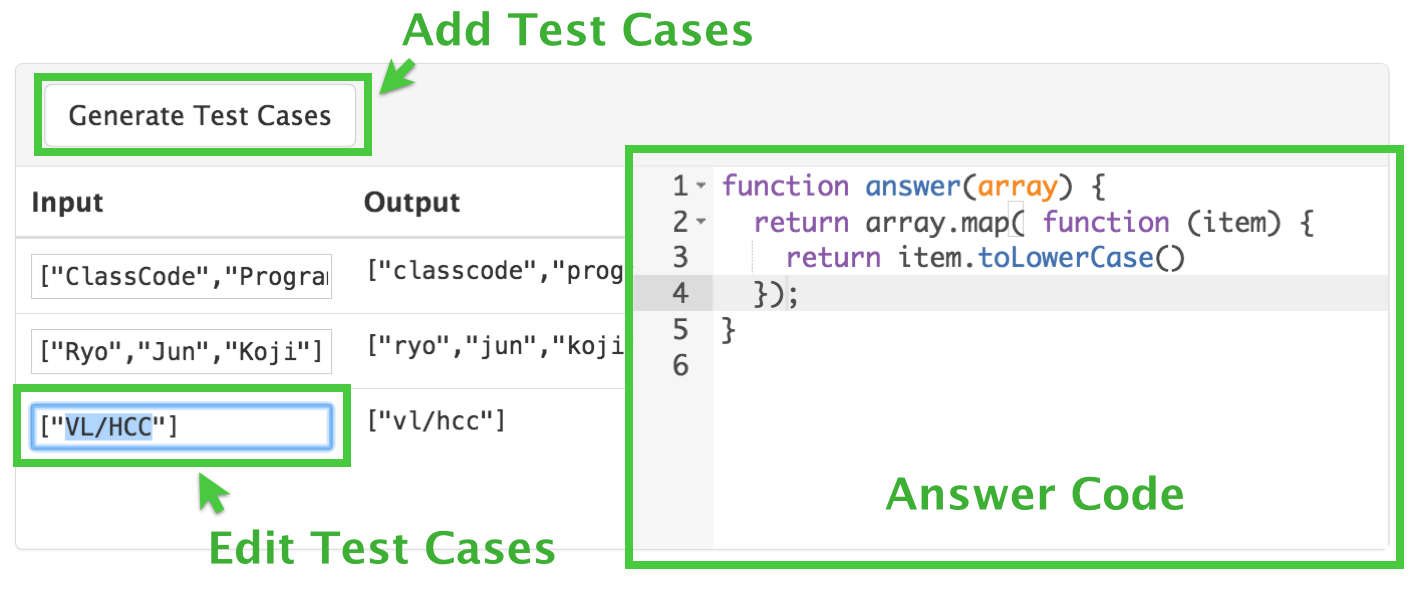}
\caption{Creating a milestone problem. Instructors can add test cases so that students can experiment their codes.}
\label{fig_7}
\end{figure}

Before the class, instructors prepare tutorials that students take in ClassCode. The interface for tutorial creation is similar to a wiki system with a Markdown format, requiring a small amount of effort on authoring (Figure \ref{fig_6}). In addition to reading materials, instructors can include quick exercises if appropriate. They just add instructions for the exercise and the answer key (i.e., one line script that they want students to write). The system performs literal matching to determine if the student's answer is correct. The system also offers a user interface for creating a milestone problem (Figure \ref{fig_7}). The instructor can add test cases with different input variables for the function students are asked to implement. The current prototype only supports four types of variables (strings, integers, arrays, and JSON objects), but this can be extended in a future system.

\begin{figure}[t]
\centering
\includegraphics[width=3.5in]{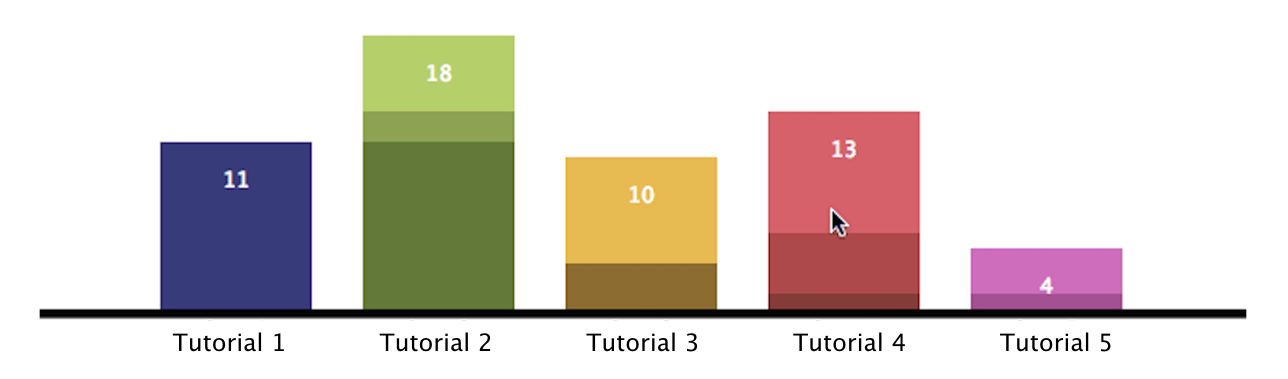}
\caption{A screenshot of a milestone problem interface. Students can write their codes and verify them with test cases prepared by instructors.}
\label{fig_8}
\end{figure}

Instructors use their tutorials as teaching materials or simply let students take by themselves. In either way, students have the freedom to go through tutorials at their own pace. ClassCode visualizes the distributions of completed tutorials as an overview of students' status (Figure \ref{fig_8}). A darker portion of the bar shows how many students are spending longer than the time threshold defined by the instructor in each tutorial. Thus, a growing darker portion indicates that a tutorial requires more time than the instructor expected. When the instructor clicks a bar in the graph, the system shows the list of the students currently taking the tutorial with their details. This list summarizes their learning activities, including the elapsed time, completed a portion of a tutorial, and results of milestone problems (e.g., the number of failures in testing).

\begin{figure}[t]
\centering
\includegraphics[width=3.5in]{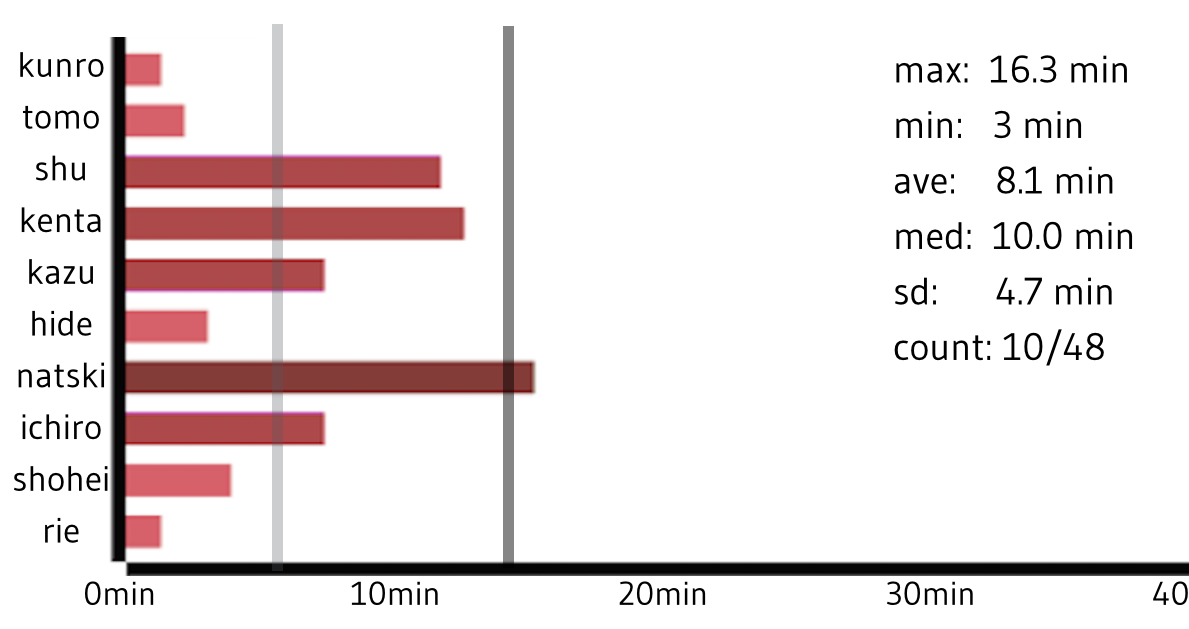}
\caption{An overview of the elapsed time of each student in the tutorial. The view also provides representative statistics of their elapsed time.}
\label{fig_9}
\end{figure}

The system also displays the amount of time each student has spent on a particular tutorial selected by instructors along with representative statistics, such as the average and standard deviation (Figure \ref{fig_9}). When the instructor clicks the graph of a student, the system shows her activity history in quick exercises and the milestone problems in a modal view. This allows the instructor to investigate how a particular student is learning and whether she is struggling with a specific issue.

ClassCode also allows the instructor to reach out to the entire group of students taking the same tutorial or its subset through online chats. After the instructor selects a student group, they can open a chat room where everyone is encouraged to post their questions and comments. This closed group communication channel enables students to discuss problems and solutions relevant to their current learning activities. The instructors may lead to online discussions or provide additional help. It is also very common that students get lost or confused by the same error or bug (e.g., a syntax error by missing a semi-colon). This chat system thus can liberate the instructor from repeatedly providing the same explanations to multiple students.

\begin{figure}[t]
\centering
\includegraphics[width=3.5in]{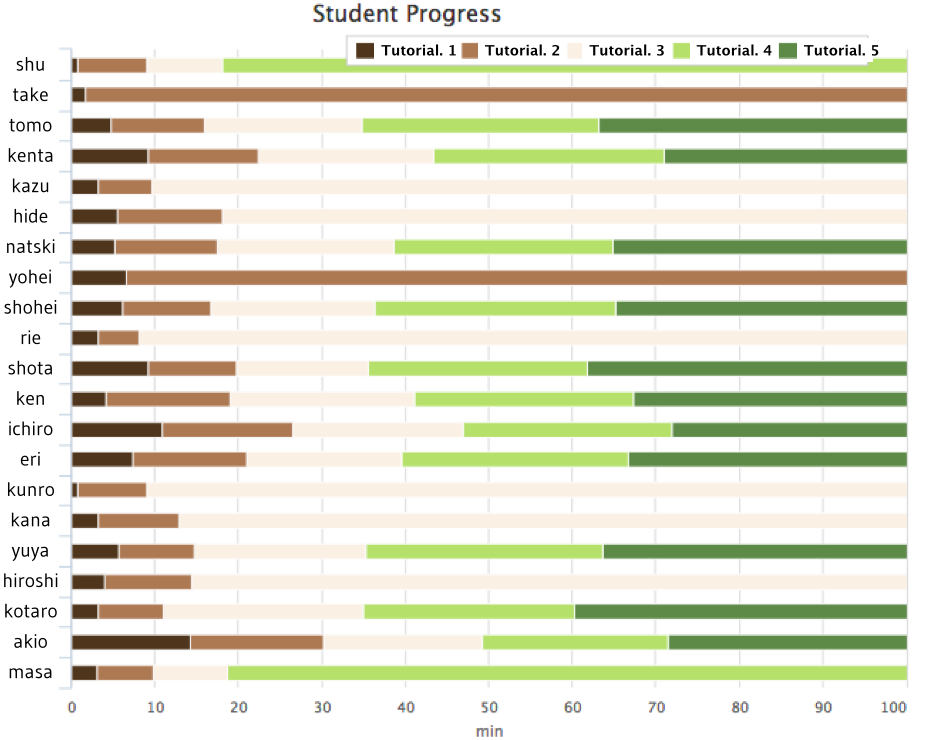}
\caption{ClassCode also summarizes the completion time of all tutorials by students in a stacked bar chart.}
\label{fig_10}
\end{figure}

\begin{figure}[t]
\centering
\includegraphics[width=3.5in]{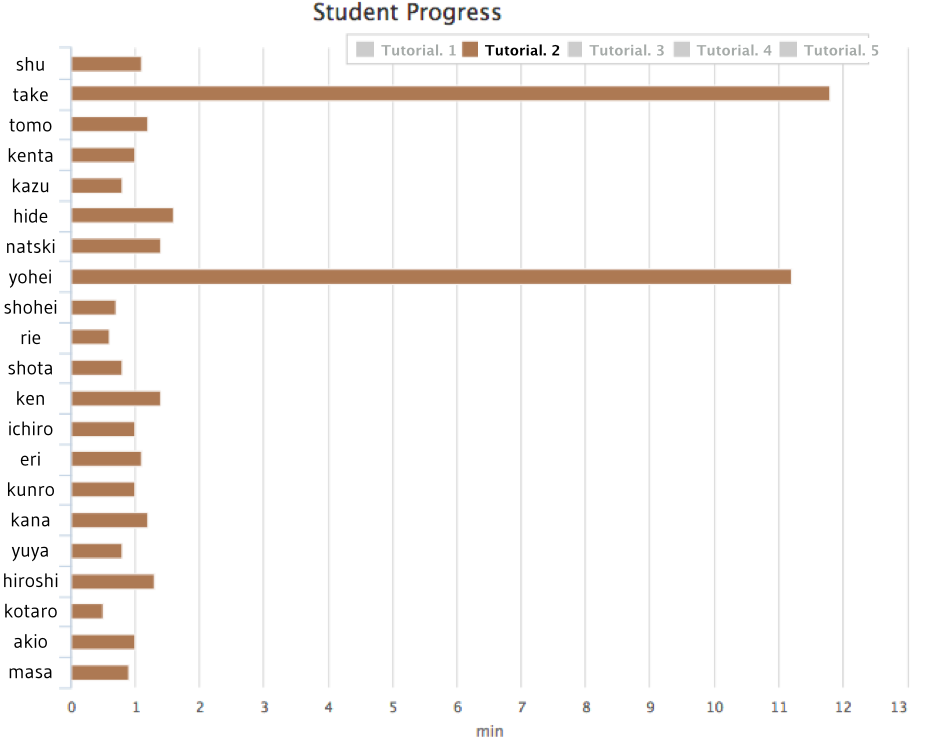}
\caption{Instructors can see the time each student spent in each tutorial.}
\label{fig_11}
\end{figure}

After the class, instructors can review the performance of students with a stacked bar visualization shown in Figure \ref{fig_10}. The length of the stacked bars represents the number of times students took to complete each tutorial. The instructor can see students' completion time for a particular tutorial by simply clicking the associated label placed on the top of the graph (Figure \ref{fig_11}). We designed this so that instructors can review their tutorials based on students' time performance.

ClassCode also supports instructors to mark codes students wrote in milestone problems. The interface shows a student's code along with marking criteria defined by the instructors. It allows markers to make annotations on the student's code and give marks by choosing the score of each criterion and then clicking the ``Mark" button. This design facilitates consistent marking by multiple markers (e.g., instructors and teaching assistants).

\subsection{System Implementation}
The current implementation of the ClassCode system uses standard web technologies: HTML5, CSS3, and JavaScript. The web server is built with Node.js (an event-driven I/O web server) and Koa.js (a JavaScript-based web application framework). The server is connected to MongoDB and Redis databases. The client-side implementation runs on a web browser and connects to the ClassCode server through the WebSocket protocol. It is used to exchange text data in real-time to enable quick exercises and bi-directional communication between an instructor and students. Data visualization is rendered on the browser with D3 and AngularJS. Although the present work only enables JavaScript programming, our system can be easily extended to support other script languages, such as Ruby and Python.

\section{User Evaluation}
We first conducted a user evaluation to assess the validity of our ClassCode interface design from the student's perspective. We recruited 9 university students (8 undergraduate and 1 graduate student; 8 male and 1 female; P1-P9) for our user study. They had some prior experience in programming and JavaScript though their skill levels were varied. They had taken at least one programming course in their universities.

Before the user study, we prepared a small course about the basics of JavaScript, covering array manipulations. The course had 5 tutorials each of which included 2.4 quick exercises on average. Due to the time constraint in the user study, we decided to make our tutorials short, which a standard learner would be able to finish in one hour.

At the beginning of the user study, participants were given an explanation about the ClassCode system and tutorials. We explained how to use and complete quick exercises and milestone problems with the first tutorial. After these explanations, the participants were asked to complete all tutorials but at the pace as they liked. One of the authors served as an instructor but did not use the chat system during the user study. All participants took approximately 1.5 hours for this part of the user study.

After the completion of tutorials, we invited the participants to a post-experiment interview. We asked their experience on ClassCode and opinions about its potential benefits and drawbacks in real classroom use. We conducted our interviews in the local language because none of the participants was English-native. We offered each participant compensation of cash equivalent to 15 USD in local currency at the end of the user study.

\section{Results}
All participants successfully completed our tutorials during the user study. We transcribed all interviews and then translated them into English as faithfully as possible. We present our exemplary quotes on participants' experience in programming courses and perceived advantages of ClassCode, highlighting potential benefits and possible improvements of the system.

\subsection{Encouraging Reviews During the Class}

Our design of quick exercises was intended to help students quickly review teaching materials and become aware of where they do not understand. Participants agreed that quick exercises helped them immediately review what they learned in tutorials. P1 appreciated the quick exercises as a way to confirm his understandings:

{\it ``If I had an exercise at the end [of a tutorial], I might just skip over where I did not understand well. But [with ClassCode], I can easily see where I did not understand."} [P1]

P2 and P5 also agreed on the benefits of quick exercises as a motivating factor of learning:

{\it ``I most clearly remembered that I was very happy when I got answers [in quick exercises] right... I enter answers and get green (an indication for correct answers), and move on to the next part. It's a fun experience, much like solving puzzle games."} [P1]

{\it ``It makes easier for novice learners to study to enter [his codes] in a console and immediately get results. It gives me a feeling of achievement."} [P5]

This set of qualitative evidence suggests that ClassCode successfully formed a closed learning loop by closely intertwining reading materials and quick exercises.

Milestone problems are designed to be another way to review students' understanding of tutorials. Our participants also positively accepted them, but for different reasons from quick exercises. The participants unanimously agreed that the ability to compare with other students' solutions was helpful. P4 found that solutions by other participants showed different perspectives and ideas and encouraged him to review his own immediately after he completed a milestone problem:

{\it ``[The solution gallery] made me aware of ideas I did not think of or solutions that were faster than my code. It also shows completely different approaches... I did not have a chance to see other students' solutions during a class. Maybe they could show me if I asked, but this system already supports it and I did not have to ask explicitly right after I finished a problem."} [P4]

P1 even thought that this solution gallery feature could be helpful for students to review tutorials at home.

{\it ``I like I can see other people's solutions. If I take tutorials by myself at home, I would try all the codes and see how they work... I would rather want to have this (the solution gallery) for reviews at home. Because I don't have such an ability in the class, and I cannot know if I have written a good code or not. I tend to think that [in the class] it’s ok as long as my code works."} [P1]

These comments imply that milestone problems can help students execute what they have learned in tutorials as well as make them aware of other solutions, stimulating further learning.

Although the current ClassCode prototype successfully demonstrated the benefits of milestone problems and solution galleries, P5 further suggested a visualization of highlighting differences between his own solution and others:

{\it ``I think it's a good idea to be able to see other students' codes, but when codes are short, they tend to be very similar, and do not help me much. If a problem is difficult enough, maybe it's helpful because other students may have written more concise codes than me... But [The solution gallery] could be less helpful when codes are too long or short. If codes are too long, people would skip reading. If they are too short, they look alike."} [P5]

A future system could employ a mechanism like OverCode \cite{glassman2014overcode} to offer a better summarization of the differences in codes.

\subsection{Lowering Entry Burdens}

Participants agreed that ClassCode could be most helpful for novice learners to start and keep learning. As ClassCode integrates both reading materials and interactive coding environments, students do not need to prepare tools for programming. Participants found that this is a strong advantage for novice learners. P2 liked ClassCode because it did not require him for a tedious setup.

{\it ``You know, it is tedious to set up the right environment when you learn to program. But [ClassCode] allows students to start learning on the Web. It's very handy and good for learners."} [P2]

P4 agreed that ClassCode removes the burden of programming environment setup that often discourages novice learners to start learning.

{\it ``When people start learning to program, I have the impression that the most difficult part is to set up a coding environment. [ClassCode] completely gets rid of it, and let people just start learning. It makes the entry really easy."} [P4]

ClassCode offers a holistic learning environment for programming. This liberates students from not only painful development environment setup but also variable and test case configurations. As a result, students can focus on learning the main topics rather than get deviated to unnecessary debugging. Our participants felt strong value on this as programming courses at schools can have many novice learners.

Another aspect of ClassCode that could support novice learners is that it allows them to learn at their own pace. P5 appreciated the freedom students can have in ClassCode:

{\it ``Because people have different levels of understandings, we always have some people who understand well and others who don't in the class. I think [ClassCode] makes learning more efficient by letting people learn as they like. If you want to review, you can review it. If you want to move on, you can move on. That's good."} [P5]

Tutorials are broken down into smaller parts in ClassCode, P4 compared ClassCode against a C language course he took before and commented on how helpful this structure was for him to digest teaching materials.

{\it ``[In the C language course he took], we were given 10 random pieces of codes and asked to sort them so that the entire code would run. Then, I tried to solve by writing code by myself, but (teaching materials) were not broken down very well. So, each step was very big. But this system (ClassCode) gave me multiple steps before reaching the final goal. It really made materials easy to understand."} [P4]

These results are indicative of our successful design of the system. Novice learners are easily left behind in programming courses in classrooms, and ClassCode has the potential to provide them with more support, which otherwise instructors would not be able to offer.

\subsection{Supporting Mutual Help between Students}

In post-experimental interviews, they expressed an unsatisfying experience in programming courses at universities. P4 answered us as follows when he could not understand teaching materials.

{\it ``No, we could not have sufficient help on getting answers to our questions because our course was large, like 100 students. So we did not have enough teaching assistants and they could not follow us up very well. If I got issues during the class, I had to review by myself or ask my friends for help after the class. But in the class, I was just very busy catching up with what the instructor was saying."} [P4]

P2 shared with us his experience of utilizing online communication channels in his programming course:

{\it ``Because we couldn't talk during the class, we used Twitter to ask each other questions. Or maybe after the class... We were new to each other at that time, so it was kinda hard to directly ask for help offline. So we posted comments on Twitter and looked for help."} [P2]

These comments suggest that students did not always have enough support during the class, resulting in an inefficient learning experience. ClassCode aims to encourage students to help each other, and our participants appreciated this approach. P5 agreed on a benefit that students can ask questions through the system and others who have bandwidth can offer help:

{\it ``I like being able to ask other students [for help]. I think students who understand can even learn more by teaching to others. And [in ClassCode], we can ask [for help] without explicitly naming who to help. It allows people who have time or know solutions to help. If I ask my friends and they don't understand either, I would take their time. I really like the system because I can post (my request for help) to the entire class and get help from those who understand."} [P5]

Our participants agreed that they would offer help to their fellow students in the class because they are in the same class. They also agreed that it would be beneficial to share thoughts and opinions among fellow students.

{\it ``I think that it is quite important to be able to teach each other. We are studying in the same class, which means that we are in the same group. I want to know what they are thinking of (course materials) and quickly ask. I think it's quite good."} [P4]

Being in one classroom can give students a feeling of belonging to the same community. The qualitative evidence above suggests that ClassCode can further encourage their reciprocity. P4 explicitly commented that solution galleries can be good learning stimuli because they can visualize how better fellow students are doing.

{\it ``I can search (solutions) on the Internet, but they are complete strangers. If I study in a class, I have some feelings about competitions or groups. Because people at more or less the same level are writing codes, I think I can compare my ability with others more realistically. If I see someone at the same level has written such a great code, it's motivating me."} [P4]

P1 also agreed to offer help to fellow students, but the size of a class could impact on his behavior. One possible reason is that students can know each other better in a smaller classroom, and may feel stronger social connections. Another explanation on this is that people like P1 might feel that their posts become public rather than shared only within a small community.

{\it ``If someone needs help and I can reply, I would do it. At my university, a class is very small... Because we have only 20 people, I would give feedback like `how about this?' even if I don't understand completely. But if I was in a larger class, I would reply to only what I am confident in."} [P1]

Another feature our participants liked in ClassCode was social feedback. P2 and P3 found that voting scores given by other participants were motivating them.

{\it ``I like that [ClassCode] has some similarities with online social networks and has a `like' button. The number of likes always gives people positive feelings in any app. It's great to have. I think it would motivate people in programming."} [P2]

{\it ``I saw a score on the top of the screen. Is this something that gets increased when I get voted, right? If I see that score going up, I feel like `Yes, good!'. That score gives me a bit of feeling of achievements."} [P3]

P2 further commented that ClassCode as a whole enables students to share their experience in real-time, making learning more engaging. These comments are indicative that ClassCode is not only supporting each student's learning but also motivating communication and mutual help among learners.

{\it ``[ClassCode] is much like social networks and I had a feeling of sharing (learning) experience. I think fun comes from experience sharing. [ClassCode] gave me such a feeling of `shared' in real-time, and I found it really important."} [P2]

\section{Expert Review}
We next conducted an expert review on ClassCode to obtain opinions from instructors. We recruited four people who had experience in teaching a programming course as an instructor (T1) and assistants (T2-T4) from our institute. Their course was an introductory programming class for undergraduate students. The class size was approximately 130-150 students. Each lecture in this course consisted of two parts. In the first half, the instructor explained some basics of C programming. Then, in the latter half, students were given exercise problems and asked to write solution codes.

We started our interviews by asking questions regarding the background and teaching experience. We then gave a brief walkthrough of ClassCode, and explain major features. Interviewees were welcome to try the system. We then asked them their impressions about ClassCode and thoughts on how it could help or interfere with their teaching.

\subsection{Results}
Our interviewees mentioned to us that they had students with a wide variety of expertise programming levels in their course.

{\it "Yeah, [students' expertise levels are] quite diverse. For example, some students do not know they have to use `==' in the if sentence (to compare variables), but some can talk about a compiler.”} [T4]

Because of such diversity, they agreed that they wanted to know how each student was doing in the classes and visualizations ClassCode offers (Figure \ref{fig_8}) could be very helpful for this purpose.

{\it ``Our instructor just oversaw students from the podium, and I don't think he could know how students were doing. So this view (a stacked-bar visualization)is really good, I think. It's probably hard for the instructor to know, for example, that students get lost because what he is saying is too difficult."} [T3]

T2 and T3 saw an additional value on the visualization to help them follow up students who seem to be confused.

{\it ``The merit is that because I can see all students' progress, so if I spot a student who is particularly slow, I can talk to her before she asks. We have some students hesitant to ask questions."} [T2]

{\it ``I think this (the visualization) is good because I can approach students who are very shy to ask though do not understand well."} [T3]

Furthermore, T3 told us that the system could be used to reach students who have completed earlier and give them additional exercises. These comments indicate that ClassCode's visualizations could potentially help instructors approach students at different skill levels more effectively.

{\it ``We have some students who feel like they (exercises) were too easy. So, we could add more exercises to these students."} [T3]

T1, an instructor, appreciated the tracking mechanism as he tried to manually do it but found that it was not very successful.

{\it ``I found this (the tracking mechanism) beneficial... In my course, I prepared eight problems for debugging. In order to know where students were, I asked them to post their progress in a separate online form, and I saw them in a spreadsheet... But students didn't submit because it was bothersome."} [T1]

Our interviewees also agreed on potential positive effects on solution galleries that students can see after the completion of milestone problems. T3 commented that the lack of capabilities to share codes may hamper mutual learning and ClassCode could resolve the issue.

{\it ``I think the reason why students do not share codes is not like they don't want to, but they don't have a platform to do it. If they have it, I think they will use it... If students feel they are still a novice, they can make their codes private, but if they have confidence, they would be willing to publish. And other students can see them and find ‘my classmates can write such great code. I should work hard.' I think it would be a great motivator."} [T3]

Quick exercises and milestone problems were also positively received by our interviewees, but T1 expressed a concern that students may use some tricks to pass test cases and do not learn what instructors expect them to understand.

{\it ``Can students see these test cases? If we do this, students might adapt their codes to these test cases... It might be problematic if students can see too much information about test cases."} [T1]

This problem can be solved by introducing intelligent analysis on test case creations like CodeHunt \cite{bishop2015code}. T1 also saw the value on the test case checking capability ClassCode offers and would like to have more complex testing, such as recursion.

{\it ``It would be good if I post a problem that requires to use recursion and the system can test if codes (submitted by students) are actually using it or not."} [T1]

\section{Discussions and Limitations}

The results of the user evaluation and expert review confirm that ClassCode can satisfy the needs of both students and instructors well. Our student participants were able to use the system quickly and saw strong values on interactive components, such as quick exercises and milestone problems. They also agreed that ClassCode could enhance communication among students through the help mechanism and solution gallery. 
For future work, this mutual learning experience would be further enhanced if the system incorporates with peer annotation and commenting systems like~\cite{suzuki2015poster}.
Our instructor participants expressed their clear agreement that ClassCode could offer improved awareness of students' learning. These results validate our design of ClassCode and show potentials in real classroom use.

Our present study has several limitations. We have not deployed our system to a real programming course at schools. Thus, it still remains unknown how ClassCode could impact on teaching and learning in classrooms. However, both our student and instructor participants expressed positive impressions with concrete ideas of how ClassCode could address issues they encountered in classes. A future deployment study will clarify how ClassCode could reshape the learning and teaching experience in classrooms.

Our current system implementation only supports JavaScript as a programming language and offers a limited capability on test case creation and checking. Feedback of test case checking to students is also only binary (i.e., passed or failed). A future system could integrate an improved way for students to investigate how errors occur in their codes. For example, Online Python Tutor \cite{guo2013online} visualizes how code is running in detail.
TraceDiff~\cite{suzuki2017exploring, suzuki2017tracediff} leverages program synthesis to generate more personalized feedback to help students understand the cause of bugs.
FindBugs \cite{ayewah2008using} can be used to help students identify bugs efficiently. We explore what improvements in interactive exercises could further improve learning efficiency in the future.

\section{Conclusion and Future Work}
Despite the increasing importance of programming education, universities and the school have not changed its teaching format, one-to-many lectures. We present ClassCode, a Web-based programming education support for classroom use. ClassCode offers an environment where students can take tutorials with interactive coding exercises intertwined at their learning pace while instructors can overview how they are learning. Our user study and expert review confirm potential merits and future improvements of ClassCode in a real classroom.

Future work will investigate the effect of ClassCode by deploying it into an actual programming course. Our current explorations focus on higher education, but we will expand it to teaching scenarios to elementary students.

\balance
\bibliographystyle{IEEEtran}
\bibliography{vl-hcc}

\end{document}